\documentclass{camera}
\usepackage{graphicx}

\begin{document}

\title{ALICE pp physics programme}

\author{Ingrid Kraus, for the ALICE collaboration}

\organization{TU Darmstadt, Schlossgartenstrasse 9, 64289 Darmstadt, Germany}

\maketitle

\begin{abstract}

The physics programme of the ALICE experiment at CERN-LHC comprises besides studies of high-energy heavy-ion collisions measurements of proton-proton interactions at unprecedented energies, too. This paper focuses on the global event characterisation in terms of the multiplicity distribution of charged hadrons and mean transverse momentum. These bulk observables become accessible because the detector features excellent track reconstruction, especially at low transverse momenta.
The measurement of strange hadrons is of particular interest since the strange-particle phase-space was found to be suppressed beyond canonical reduction at lower center-of-mass energies and the production mechanism of soft particles is not yet fully understood.
Here we benefit in particular from particle identification down to very low transverse momentum, i.e. 100 - 300 MeV/c, giving access to spectra and integrated yields of identified hadrons. \\
Equipped with these features, ALICE will play a complementary role w.r.t. other LHC experiments. New interest in the soft part of pp collisions arose recently and new insights in the physics of the underlying event are expected from both, theory and experiment.

\end{abstract}

\section{Introduction} 
\begin{figure}
\includegraphics[width=0.90\linewidth]{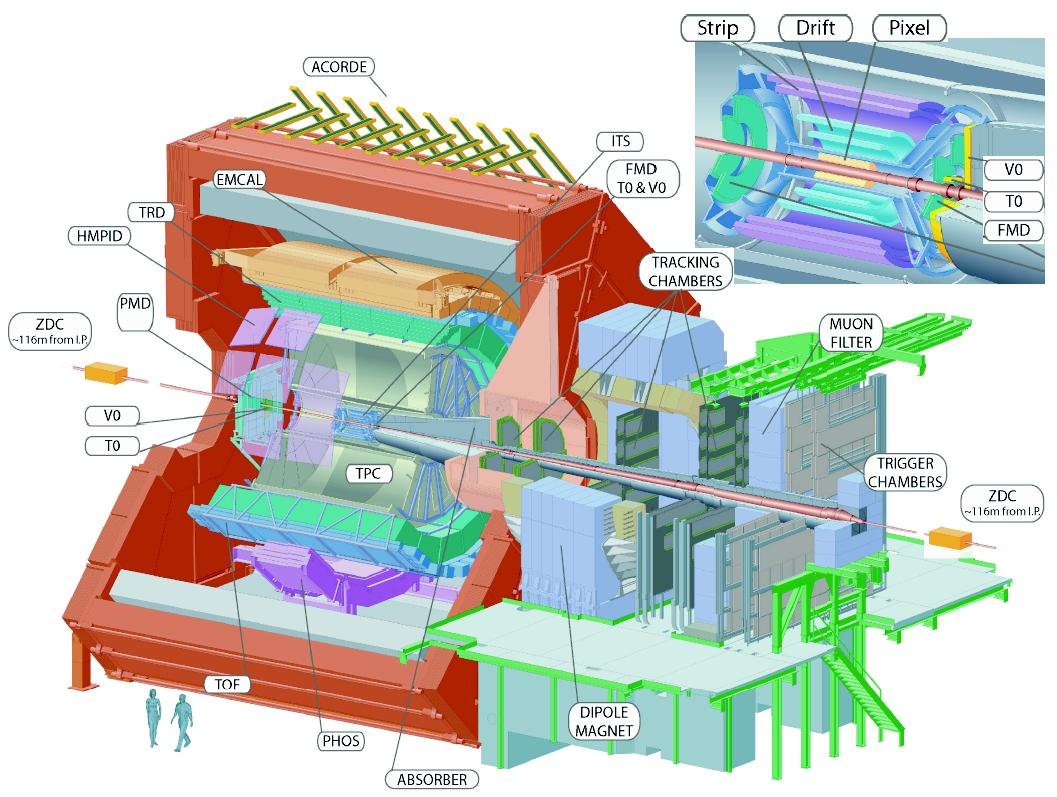}
\caption{The ALICE detectors.}
\label{fig1}
\end{figure}
ALICE (A Large Ion Collider Experiment) is the dedicated heavy-ion experiment at the Large Hadron Collider (LHC) in Geneva. The search for deconfined nuclear matter is presented in another article in these proceedings. Here, we outline how ALICE can contribute to the proton-proton programme. 

The design aims at measurements down to low transverse momentum, $p_t$, therefore a moderate magnetic field of up to 0.5T is provided by the L3 magnet. Inside the solenoid, the central-barrel detectors are arranged in a typical onion-shape, Fig.~\ref{fig1}. The inner tracking system, ITS, consists of 3 different silicon technologies (pixel, drift, strip), 2 layers each. It is surrounded by the time projection chamber, TPC. With up to 160 space points it is not only the main tracking detector but allows to identify particles by specific energy loss in the gas. A particle leaving the TPC has crossed only 10\% of a radiation length after traversing 2.6~m.
First analyses of event multiplicity and (identified) hadron spectra rely mainly on the ITS and TPC detector systems. To study global event properties, a few 10,000 to a few 100,000 events are sufficient which corresponds to some hours of data taking.

ALICE will make further strong contribution to the LHC pp programme, such as studies of baryon number transport and femtoscopy as well as baryon/meson production at intermediate $p_t$, jet reconstruction and charm cross section measurements. Heavy flavour analyses in general profit by the reconstruction of decay vertices thanks to the high spatial resolution of the ITS.
Measurements at higher $p_t$, electron identification and photon studies benefit also from outer detector components, namely the transition radiation detector, TRD, the time of flight system, TOF, and smaller coverage cherenkov, HMPID, and photon, PHOS, detectors and a calorimeter, EMCAL. Muons are measured in the forward arm in tracking and triggering stations following the absorbers.
\section{Multiplicity distribution} 
\begin{figure}
\includegraphics[width=0.4\linewidth]{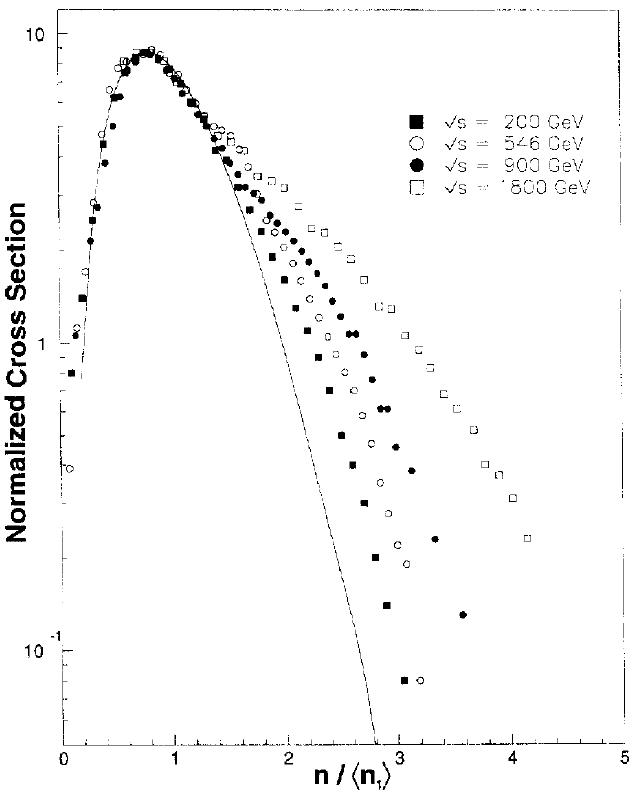}
\includegraphics[width=0.6\linewidth]{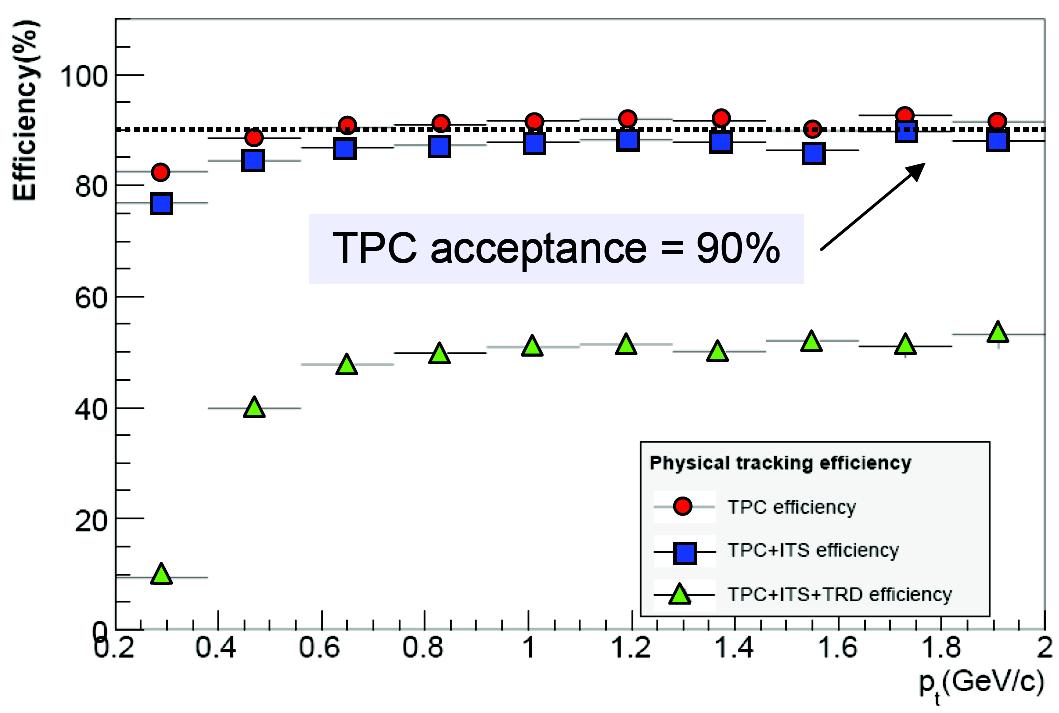}
\caption{Left: Multiplicity distributions measured various c.m. energies ranging from 200 GeV to 1800 GeV by the UA5 and E735 experiments at SppS and Tevatron resp.~\cite{e735}. The full line is a parametrisation of ISR data.
Right: Tracking efficiency as a function of transverse momentum for TPC stand-alone tracking and combinations with other detectors in ALICE.}
\label{fig2}
\end{figure}
The KNO scaling~\cite{kno}, i.e. the fact the normalised multiplicity distribution keeps its form independently of the beam energy and just scales up as $\ln(s)$, was proposed in the early 70's and confirmed in the entire ISR energy range. However, at SppS it did not hold anymore and detailed studies at the Tevatron revealed an extra structure in the multiplicity distribution compared to ISR data, Fig~\ref{fig2}. The assumption that this shoulder reflects an emerging second or multiple parton-parton interaction can be tested at LHC by measuring the cross section of this process as it should increase with c.m. energy and, possibly, an arising third semi-hard collision.

The analysis makes use of the silicon pixel detector, SPD, and TPC, both exhibit full azimuthal coverage in the central rapidity region, $|\eta|<$0.9. The tracking efficiency is of 80\% for $p_t$ = 0.3 GeV/c tracks, Fig~\ref{fig2}, and rises quickly to 90\% which is the limit dictated by geometrical acceptance. Low momentum particles are strongly bent in the magnetic field. This leads to a higher efficiency in the SPD since it is closer to the interaction point ($r_{\rm SPD}$ = 7~cm, $r_{\rm TPC} \approx$ 165~cm for a track with at least 50 clusters) and increases the momentum reach towards very soft particles. The SPD misses less than 1\% of the spectrum at very low $p_t$.
\section{Mean transverse momentum} 
\begin{figure}
\includegraphics[width=0.45\linewidth]{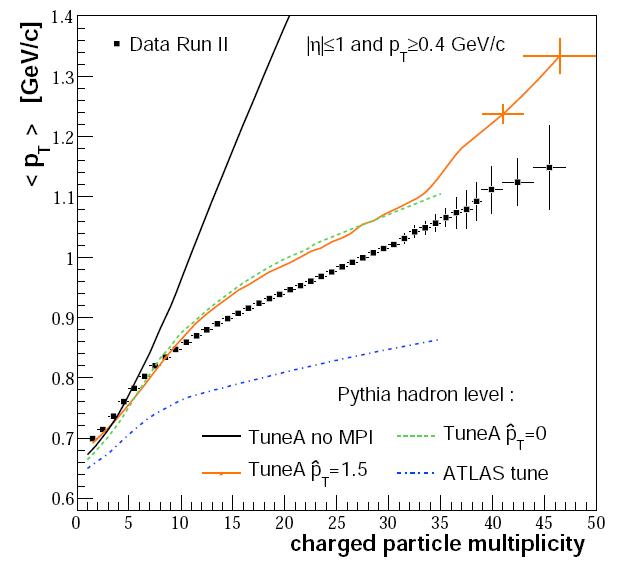}
\includegraphics[width=0.55\linewidth]{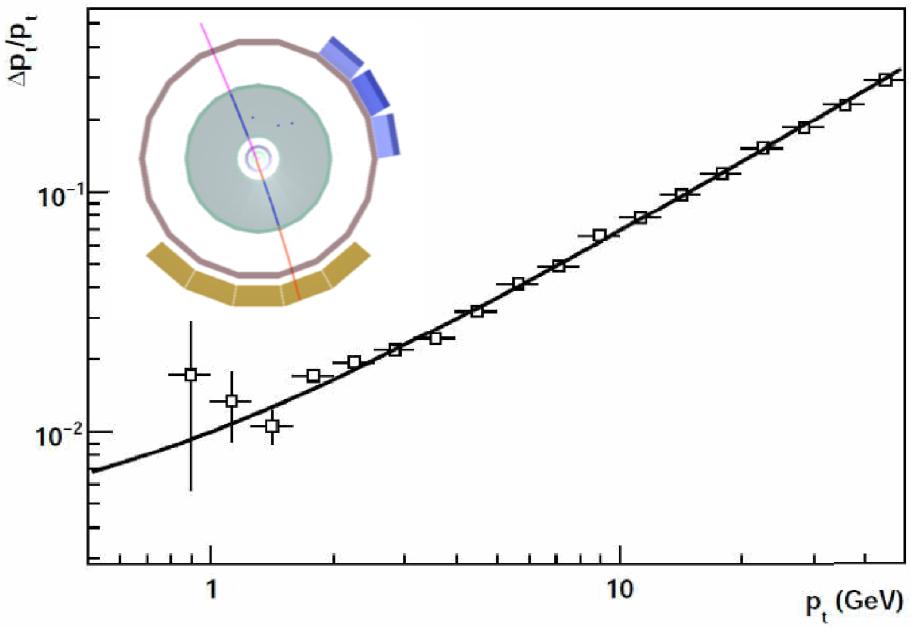}
\caption{Left: Mean transverse momentum as a function of multiplicity measured by the CDF experiment at highest Tevatron energy of 1.96 TeV~\cite{cdf}, together with Pythia calculations.
Right: Transverse momentum resolution $\Delta  p_t / p_t$ in the TPC (before calibration) as a function of transverse momentum $p_t$. The insert shows an example of a cosmic track used to derive the momentum resolution (by comparing the upper and lower half).}
\label{fig3}
\end{figure}
In the early 80's van Hove suggested to measure the correlation between mean $p_t$, $\langle p_t \rangle$, and multiplicity~\cite{hove}. Using $\langle p_t \rangle$ as a thermometer and the charged-particle multiplicity $N_{ch}$ as a measure for the entropy density, he proposed a second rise after the flattening as a signature for the deconfined phase. The obvious absence, Fig.~\ref{fig3}, could be due to a lack of available energy and the rise might be observable at LHC. Models like Herwig and Pythia failed to explain the data unless multi-parton interactions, MPI, are included (see Ref.~{\cite{cdf}}). A couple of parton-parton interactions in parallel can intelligibly explain events with high multiplicity and moderate $\langle p_t \rangle$.

ALICE will measure the excitation function of this correlation, which links bulk production at low $N_{ch}$ with the jet-fragmentation dominated regime at high $N_{ch}$. The $p_t$ resolution of TPC tracks, measured using cosmic triggers, Fig.~\ref{fig3}, is better than 5\% for tracks with $p_t ~ <$ 5 GeV/c relevant for early publications. The design value of 5\% at $p_t$ = 10 GeV/c~\cite{ppr} will be reached when the more elaborate calibrations are completed. Not only the cosmic particles are used but also data created with a laser system illuminating the detector gas. Adding the interaction vertex will further improve the precision.
\section{Strangeness production} 
\begin{figure}
\includegraphics[width=0.45\linewidth]{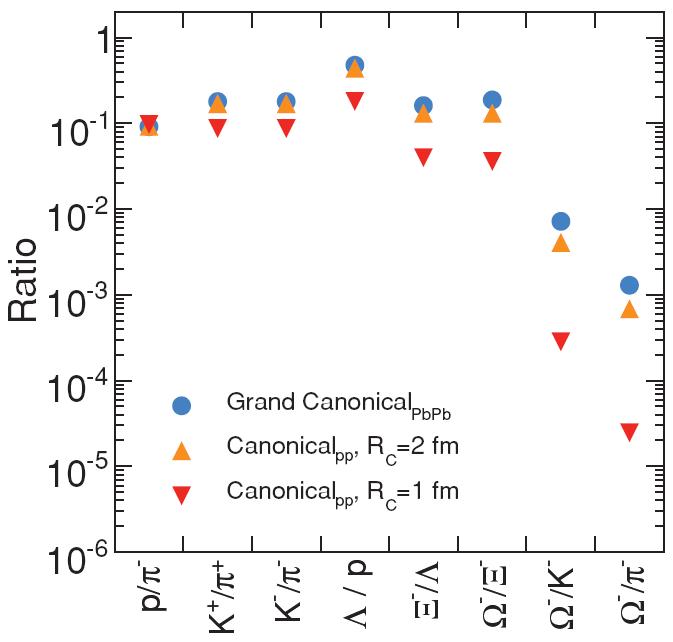}
\includegraphics[width=0.55\linewidth]{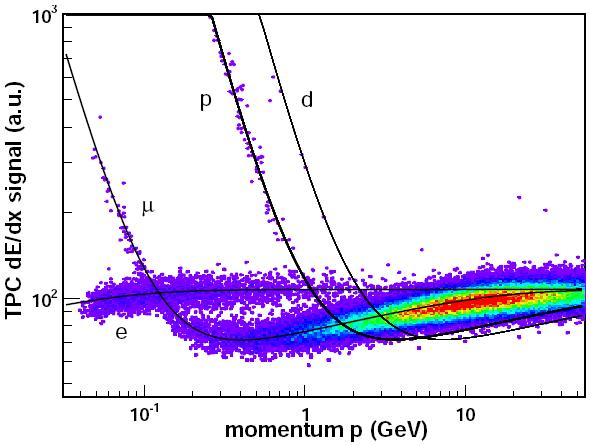}
\caption{Left: Statistical model predictions for various particle ratios in pp collisions at LHC, using different values for the cluster size $R_C$~\cite{stmo}. Right: Differential energy loss $dE/dx$ of charged tracks in the TPC measured with cosmic trigger as a function of momenta, together with Bethe-Bloch curves for different particle species.}
\label{fig4}
\end{figure}
Application of the statistical model to particle production in heavy-ion and, particularly, in elementary particle collisions revealed that canonical suppression alone is not sufficient to quantify the observed strange-particle yields~\cite{gams}. Clusters with local strangeness conservation inside the fireball provide stronger reduction of the strange-particle phase-space~\cite{stmo}. Studies of data at SPS and RHIC energies resulted in correlation lenght' of the order of 1~fm. Its extrapolation to higher energies is ambiguous. The same cluster size at LHC is expected if strangeness production is governed at early time by e.g. the maximal energy density. Increasing subvolumes point to a late stage with an expanded system with many produced particles. Especially multi-strange baryons are sensitive to the correlation length $R_C$, Fig.~\ref{fig4}.

\begin{figure}
\includegraphics[width=0.8\linewidth]{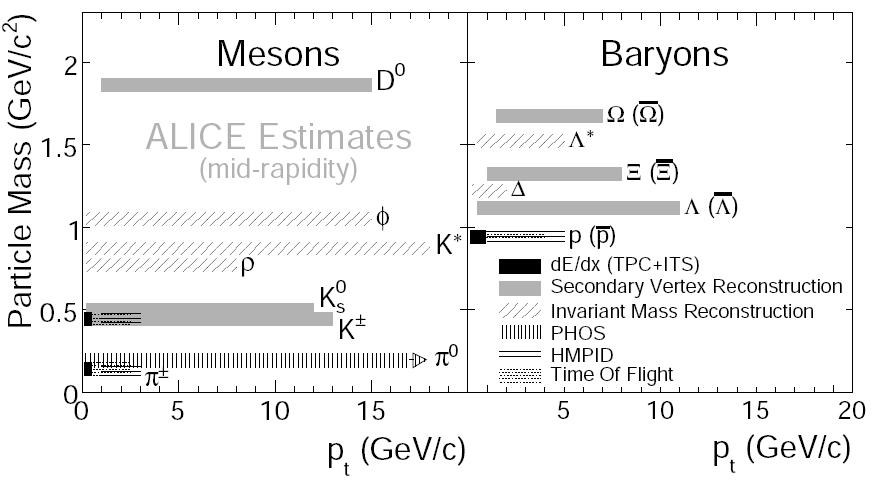}
\caption{Transverse momentum range of identification of mesons and baryons at mid-rapidity.}
\label{fig5}
\end{figure}

Particles are identified in the TPC by their specific energy loss dE/dx. A resolution of 5.7~\% for particles from cosmic trigger events is achieved after calibration with radioactive Krypton, Fig.~\ref{fig4}. The limited momentum range ($p ~ \leq$ 1 GeV/c) is extended by exploiting other detectors and techniques, Fig.~\ref{fig5}. TOF and HMPID allow to identify particles with a few GeV/c. Invariant mass reconstruction without (for short living resonances) and with secondary vertex reconstruction (for weak decays of strange hadrons and heavy flavour carrying particles) will be feasible over the entire range and is essentially statistics driven.
\section{Summary} 
A rich physics programme, addressed in pp reactions, has been developed in ALICE.
It focusses mainly on the bulk of particle production at low transverse momentum.
These analyses are feasible thanks to large acceptance in azimuth, efficiency down to very low $p_t$, and good resolution in both momentum and particle identification. We await eagerly first collisions.
\end{document}